\documentclass[letterpaper, 10 pt, conference]{ieeeconf}  

\IEEEoverridecommandlockouts                              

\usepackage{graphics} 
\usepackage{epsfig} 
\usepackage{amsmath} 
\usepackage{amssymb}  
\usepackage{algorithm}
\usepackage{algpseudocode}
\usepackage{mathtools}
\usepackage{graphicx,subfigure}
\usepackage[usenames,dvipsnames]{color}
\usepackage[nospace]{cite}

\newcommand{\HLB}[1]{{\color{blue} #1} \color{black}}
\renewcommand{\HLB}[1]{ #1}

\newcommand{\absmin}{\operatornamewithlimits{absmin}}

\newtheorem{corollary}{\bf{Corollary}}

\title{\LARGE \bf A Topology-Guided Path Integral Approach \\ for Stochastic Optimal Control}

\author{Jung-Su Ha and Han-Lim Choi
\thanks{J.-S. Ha is with the \HLB{Department} of Aerospace Engineering, KAIST, Daejeon, Korea       {\tt\small wjdtn1404@kaist.ac.kr}}%
\thanks{H.-L. Choi is with the \HLB{Department} of Aerospace Engineering \& the Center of Field Robotics for Innovation, Exploration, and Defense, KAIST, Daejeon, Korea       {\tt\small hanlimc@kaist.ac.kr}}
}

\begin{document}

\maketitle
\thispagestyle{empty}
\pagestyle{empty}

\begin{abstract}
This work presents an efficient method to solve a class of continuous-time, continuous-space stochastic optimal control problems of robot motion in a cluttered environment.
The method builds upon a path integral representation of the stochastic optimal control problem that allows computation of the optimal solution through sampling and estimation process.
As this sampling process often leads to a local minimum especially when the state space is highly non-convex due to the obstacle field, we present an efficient method to alleviate this issue by devising a proposed topological motion planning algorithm.
Combined with a receding-horizon scheme in execution of the optimal control solution, the proposed method can generate a dynamically feasible and collision-free trajectory while reducing concern about local optima.
Illustrative numerical examples are presented to demonstrate the applicability and validity of the proposed approach.

\end{abstract}

\section{Introduction}
Computing the optimal policy for a system driven by some uncertain disturbance, which is called a \textit{stochastic optimal control problem}, is one of the most important problems in planning/control of robotic platforms in a complex environment.
In a discrete-time/discrete-state and control space setting, the problem is formulated as a Markov decision process (MDP) and solved through the dynamic programming procedure, e.g. value iteration or policy iteration.
The problem in a continuous setting, which is of the primary interest of this work, can be solved in a similar manner if transformed into a discretized version; however, this discretization approach is not scalable for a high-dimensional state space.
Alternatively, an optimality condition for the continuous problem itself can be derived and utilized. It is well known that the optimality condition results in a nonlinear partial differential equation (PDE), called the Hamilton-Jacobi-Bellman equation; but, solving a nonlinear PDE is intractable in most robotic control applications.

Fortunately, there is a class of stochastic optimal control problem, called linearly-solvable optimal control (LSOC)~\cite{todorov2009efficient}, for which the HJB equation can be solved in a more efficient way with appropriate reformulation.
For an LSOC problem, the notion of a \textit{desirability} function, which is effectively an exponential value function, is introduced in order to transcribe the original nonlinear HJB equation on the value function into a linear PDE on the desirability function. In addition, it has been found that the Feynman-Kac formula allows the solution of such linear PDE to be expressed as an expectation of some path integral.
As a result, the stochastic optimal control problem is transformed into an estimation problem, which can be solved by sampling a set of stochastic paths and then evaluating their expectation.
This aforementioned procedure to solve a LSOC is referred to as path integral (PI) control~\cite{kappen2005path}.
For more interesting views and different derivations of PI control, we would refer the reader to \cite{theodorou2015nonlinear} and references therein.

Advanced estimation techniques, such as importance sampling, can be applied to effectively solve the aforementioned transformed  problem of a LSOC.
In \cite{theodorou2010generalized,theodorou2010reinforcement}, the control policy is parameterized and then estimated using an importance sampling technique on the basis of the path integral formula.
In \cite{thijssen2015path}, path-integral formula is utilized to construct a state-dependent feedback controller and theoretical analysis on how sampling strategies affect the estimation results is presented.
In \cite{kappen2015adaptive}, the cross entropy method was applied to build an efficient importance sampler that reduces estimation variance.
In \cite{arslan2014information}, the rapidly-exploring random tree (RRT) algorithm was used to help the importance sampler to pick valuable samples.

This paper addresses a continuous LSOC problem, especially in a complex configuration space with obstacles, in the path integral control framework.
This type of problem may have many local optima, since  the state space is often highly non-convex due to obstacle regions.
Thus, a sampler for PI control needs to be able to generate samples diverse and spread enough in order not to be trapped into a local minimum; however, it is not particularly easy for many conventional sampling schemes to generate samples very far from most of other samples.
To tackle this issue, the approach in this work is, therefore, (i) first specifies all possible local minima caused by obstacles for deterministic approximation problem and then (ii) generates samples around all these local minima taking them as \textit{reference trajectories}.
If the global minimum of the original problem is near one of these \textit{references}, this way eventually results in finding the global optimal solution.

Specifically in the context of motion planning in a cluttered environment, each local minimum can be associated with a different topological class; thus, a motion planner that can produce a optimized motion trajectory for every different topological class is required to support the above two-step process. There have been some attempts to build a topology-embedded path planner (although not in the context of stochastic control).
In \cite{bhattacharya2012topological,bhattacharya2013invariants}, the authors introduced/proposed the concept of $H$-signature to distinguish different homology classes of trajectories and incorporate it into a graph search algorithm.
This concept of $H$-signature is valid, but it is known to be difficult for the graph search algorithm to handle high-dimensional state space and system dynamics.
Therefore, this work proposes to incorporate the $H$-signature concept into a sampling-based motion planner that can handle more diverse class of motion planning problems with system dynamics.

Consequently, this paper presents an algorithm, termed path-integral with rapidly-exploring random homology-embedded tree star (PI-RRHT*), that consists of a homology-embedded optimal motion planner to identify all the local minima of the deterministic approximation problem and an importance sampler that solves a transformed estimation problem of the original LSOC. Combined with a receding-horizon scheme for plan \& execution of the stochastic optimal solution, the proposed method can produces the globally optimal, dynamically feasible collision-free trajectory for stochastic systems.

The procedure expanding trees to the \textit{topological concept augmented space} is reminiscent of that of the rapidly-exploring random belief tree (RRBT) \cite{bry2011rapidly} algorithm, which expands a graph in the state space using the rapidly-exploring random graph (RRG) \cite{karaman2011sampling} algorithm and projects an associated tree onto the \textit{belief space}.
Also, the proposed method in this work takes advantage of the architecture of the goal-rooted feedback motion Trees (GR-FMTs) \cite{jeon2015optimal} algorithm: it first expands a goal-rooted backward tree over a topology-embedded state space without considering uncertainty in the dynamics. Then, instead of directly using the tree as a feedback policy, the control policy for LSCO is computed using the path integral control method.

\section{Problem Description}
Let a state space, $\chi$, and control input space, $U$, be a compact subset of $\mathbb{R}^n$ and $\mathbb{R}^m$, respectively.
An obstacle region, $\chi_{obs}$, and a goal region, $\chi_{goal}$, are compact subsets of $\chi$.
Then a domain of the problem is defined by $D\subset \text{cl}(\chi\setminus(\chi_{obs}\cup\chi_{goal}))$, which is the closure of its interior, $D^\circ$ and has a smooth boundary $\partial D$;
the boundary of goal region is denoted by $\partial D_{goal}$.

Suppose $\mathbf{x}\in D$ and $\mathbf{u}\in U$ are a state and control vector, respectively, $\mathbf{w}$ is an $m$-dimensional zero-mean Brownian motion process.
Consider the stochastic dynamics of which deterministic drift term is affine in control input:
\begin{equation}
d\mathbf{x} =  \mathbf{f}(\mathbf{x})dt+ G(\mathbf{x})\mathbf{u}dt + B(\mathbf{x})d\mathbf{w} \label{eq:dyn_control}
\end{equation}
where $\mathbf{f}:D\rightarrow \mathbb{R}^n$ is the passive dynamics and $G:D\rightarrow \mathbb{R}^{n\times m}$ is control transition matrix and $B:D\rightarrow \mathbb{R}^{n\times m}$ is the diffusion matrix function.
In this work, the state is assumed to be partitioned as $\mathbf{x}=[\mathbf{x}_m'~\mathbf{x}_c']'$ and then other terms are partitioned as $\mathbf{f}(\mathbf{x}) = [\mathbf{f}_m(\mathbf{x})'~\mathbf{f}_c(\mathbf{x})']'$, $G(\mathbf{x}) = [\mathbf{0}_{(n-m)\times m}'~G_c(\mathbf{x})']'$ and $B(\mathbf{x}) = [\mathbf{0}_{(n-m)\times m}'~B_c(\mathbf{x})']'$.\footnote{The prime sign, $(\cdot)'$, throughout the paper denotes the transpose of a matrix.}
It is also assumed that $G_c:D\rightarrow \mathbb{R}^{m\times m}$ and $B_c:D\rightarrow \mathbb{R}^{m\times m}$ are invertible.

The objective of the problem is to find a control policy which achieves the goal region while avoiding collision with other boundaries (e.g. obstacles) and also minimizes the cost functional.
The problem is formulated as a first-exit stochastic optimal control problem.
Let a function $q: D\times U \rightarrow \mathbb{\bar{R}}$ and $\phi:\partial D \rightarrow \mathbb{\bar{R}}$ be an instantaneous state cost rate and a final cost function, respectively, where $\mathbb{\bar{R}}$ denotes the extended real number line $\mathbb{R}\cup\{-\infty,+\infty\}$.
For given control policy $\pi:D\rightarrow U$, the cost functional which we want to minimize is defined as:
\begin{equation}
J^\mathbf{\pi}(\mathbf{x}) = E\left[\phi(\mathbf{x}(t_f))+\int^{t_f}_t q(\mathbf{x})+\frac{1}{2}\mathbf{u}'R(\mathbf{x})\mathbf{u}d\tau\right],
\end{equation}
where the \textit{first-exit time} $t_f$ is the time when the resulting trajectory first reaches the boundary of the domain, i.e.
\begin{equation}
t_f \equiv \inf\{t\geq 0: \mathbf{x}(t) \notin D^\circ\}, \label{eq:firstexit}
\end{equation}
where $\mathbf{x}(t)$ is a solution of (\ref{eq:dyn_control}) under the control policy $\pi$.
The final cost function penalizes the situation the trajectory cannot reaches the goal boundary, $\partial D_{goal}$.

\section{Linearly-Solvable Stochastic Optimal Control}
\subsection{Path Integral Control}
The optimal cost-to-go function is defined as:
\begin{equation}
v(\mathbf{x}) \equiv \inf_\pi J^\pi(\mathbf{x}),
\end{equation}
and the associated Hamilton-Jacobi-Bellman (HJB) equation is given by:
\begin{align}
0 = \min_\mathbf{u}(q+\frac{1}{2}\mathbf{u}'R\mathbf{u}+(f+G\mathbf{u})'v_\mathbf{x}+\frac{1}{2}\text{tr}(BB'v_{\mathbf{xx}})). \label{eq:HJB2}
\end{align}
From the HJB equation, the optimal control law is obtained analytically as:
\begin{equation}
\mathbf{u}^*(\mathbf{x}) = -R(\mathbf{x})^{-1}G(\mathbf{x})'v_\mathbf{x}(\mathbf{x}).\label{eq:opt_cont}
\end{equation}
Substituting this optimal control law to (\ref{eq:HJB2}) yields the second order nonlinear partial differential equation (PDE):
\begin{align}
0 = q(\mathbf{x})+v_\mathbf{x}'f(\mathbf{x})-\frac{1}{2}v_\mathbf{x}'G(\mathbf{x})R(\mathbf{x})^{-1}G(\mathbf{x})'v_\mathbf{x} \nonumber\\
+\frac{1}{2}\text{tr}(v_{\mathbf{xx}}B(\mathbf{x})B'(\mathbf{x})).\label{eq:non_PDE}
\end{align}

Due to its nonlinearity, solving the above PDE is intractable.
The nonlinearity can be removed by introducing the \textit{desirability function}:
\begin{equation}
\psi(\mathbf{x}) = \exp(-\frac{1}{\lambda}v(\mathbf{x})),
\end{equation}
where a scalar, $\lambda$ comes from the relation,
\begin{equation}
\lambda G(\mathbf{x})R(\mathbf{x})^{-1}G(\mathbf{x})' = B(\mathbf{x})B(\mathbf{x})'.
\end{equation}
This restriction means that the control and noise affect the dynamics on the same subspace and in the same direction and the control cost is reversely related to the noise scale \cite{kappen2005path,todorov2009efficient}.
Roughly speaking, with the above restriction the control is more expensive for the direction that the noise is smaller.
Rewriting the PDE in (\ref{eq:non_PDE}) with respect to $\psi(\mathbf{x})$ induces the second order linear PDE as:
\begin{equation}
0 = -\frac{1}{\lambda}q(\mathbf{x})\psi+f(\mathbf{x})'\psi_\mathbf{x}+\frac{1}{2}\text{tr}(\psi_{\mathbf{xx}} B(\mathbf{x})B(\mathbf{x})'), \mathbf{x}\in D^\circ, \label{eq:lin_PDE}
\end{equation}
where the boundary condition is given by:
\begin{equation}
\psi(\mathbf{x}) = \exp(-\frac{1}{\lambda}\phi(\mathbf{x})), \mathbf{x}\in \partial D. \label{eq:lin_PDE_bnd}
\end{equation}

The problem in (\ref{eq:lin_PDE}) and (\ref{eq:lin_PDE_bnd}) is called the Dirichlet problem associated with an elliptic operator\cite{karatzas2012brownian}.
The solution can be represented probabilistically by the Feynman-Kac formula.
Following corollary is directly modified from Proposition 7.2 in \cite{karatzas2012brownian}.
\begin{corollary}[Feynman-Kac]
Let $\mathbf{x}(t)$ be a solution of
\begin{equation}
d\mathbf{x} = f(\mathbf{x})dt + B(\mathbf{x})d\mathbf{w}^{(0)} \label{eq:dyn_passive}
\end{equation} and $t_f$ be a first-exit time as (\ref{eq:firstexit}).
If $E_P[t_f]<\infty,\forall \mathbf{x}\in D$ then, a solution of the Dirichlet problem (\ref{eq:lin_PDE}) and (\ref{eq:lin_PDE_bnd}) is given by:
\begin{equation}
\psi(\mathbf{x}) = E_P\left[\psi(\mathbf{x}(t_f))\exp\left(-\frac{1}{\lambda}\int^{t_f}_tq(\mathbf{x})d\tau\right)\right], \label{eq:desir_fun}
\end{equation}
where the expectation $E_P[\cdot]$ is taken over all trajectories $\mathbf{x}(t),~t\in[0,t_f]$.
\end{corollary}

The optimal control (\ref{eq:opt_cont}) is written with respect to $\psi$ as:
\begin{align}
\mathbf{u}^*(\mathbf{x}) &= \lambda R(\mathbf{x})^{-1}G(\mathbf{x})'\frac{\psi_\mathbf{x}(\mathbf{x})}{\psi(\mathbf{x})}\nonumber\\
& = \lambda R(\mathbf{x})^{-1}G_c(\mathbf{x})'\frac{\psi_\mathbf{x_c}(\mathbf{x})}{\psi(\mathbf{x})}.\label{eq:opt_cont2}
\end{align}
Equation (\ref{eq:desir_fun}) can be expressed as
\begin{equation}
\psi(\mathbf{x}) = \int W(\vec{\mathbf{x}}) P(\vec{\mathbf{x}})d\vec{\mathbf{x}}, \label{eq:est_cost}
\end{equation}
where $W(\vec{\mathbf{x}}) = \psi(\mathbf{x}(t_f))\exp\left(-\frac{1}{\lambda}\int^{t_f}_tq(\mathbf{x(\tau)})d\tau\right)$ and $\vec{\mathbf{x}}$ and $P(\vec{\mathbf{x}})$ represent trajectories and its probability measure, respectively.
From the path integral formulation~\cite{kappen2005path}, the probability measure of trajectory is given by:
\begin{equation}
P(\vec{\mathbf{x}}) = c\lim_{dt\rightarrow 0}\exp\left(-\frac{1}{2\lambda}\sum_{j=1}^{N} \left[ \left\| \mu(\mathbf{x}_j) \right\|^2 _{\Sigma_c(\mathbf{x}(t_j))^{-1}}\right] dt\right),
\end{equation}
where $t_1 = t,~t_N = t_f$, $\Sigma_c(\mathbf{x}) = G(\mathbf{x})R(\mathbf{x})^{-1}G(\mathbf{x})' =B_c(\mathbf{x})B_c(\mathbf{x})'/\lambda$ and $\mu(\mathbf{x}_j)\equiv \frac{\mathbf{x}_c(t_j+dt)-\mathbf{x}_c(t_j)}{dt}-\mathbf{f}_c(\mathbf{x}(t_j))$ and $c$ is a normalization constant for $\int dP(\vec{\mathbf{x}})=1$.
Partial derivative of $P$ is given by:
\begin{equation}
\frac{\partial}{\partial \mathbf{x}_c(t_1)}P(\vec{\mathbf{x}}) = \frac{1}{\lambda}\mu(\mathbf{x}_1)'\Sigma_c(\mathbf{x}(t_1))^{-1}P(\vec{\mathbf{x}}),
\end{equation}
which yields
\begin{align}
\psi_{\mathbf{x}_c}(\mathbf{x}) &= \frac{1}{\lambda}\int W(\vec{\mathbf{x}})\Sigma_c^{-1}(\mathbf{x})\mu(\mathbf{x}_1)P(\vec{\mathbf{x}})d\vec{\mathbf{x}}, \nonumber\\
&= \frac{1}{\lambda}E_P\left[W(\vec{\mathbf{x}})\Sigma_c^{-1}(\mathbf{x})\mu(\mathbf{x}_1)\right].
\end{align}
The optimal control (\ref{eq:opt_cont2}) is expressed as
\begin{align}
\mathbf{u}^*(\mathbf{x})dt &= \frac{1}{\psi(\mathbf{x})}R(\mathbf{x})^{-1}G_c(\mathbf{x})'\Sigma_c(\mathbf{x})^{-1}E_P\left[W(\vec{\mathbf{x}})\mu(\mathbf{x})dt\right],\nonumber\\
&=\frac{1}{\psi(\mathbf{x})}G_c^{-1}(\mathbf{x})B_c(\mathbf{x})E_P\left[W(\vec{\mathbf{x}})d\mathbf{w}^{(0)}\right], \label{eq:est_cont}
\end{align}
using $\mu(\mathbf{x})dt = B_c(\mathbf{x})d\mathbf{w}^{(0)}$ and $R(\mathbf{x})^{-1}G_c(\mathbf{x})'\Sigma_c(\mathbf{x})^{-1} = G_c^{-1}(\mathbf{x})$.

The desirability function and the optimal control can be estimated from Monte-Carlo (MC) sampling procedure; the estimations for state $\mathbf{x}$ with $N$ sample trajectories are given by
\begin{equation}
\hat{\psi}(\mathbf{x}) = \frac{1}{N}\sum_{k=1}^Nw^k, \label{eq:est_cost2}
\end{equation}
and
\begin{equation}
\hat{\mathbf{u}}(\mathbf{x})\delta t = \frac{1}{N\hat{\psi}(\mathbf{x})}G_c^{-1}(\mathbf{x})B_c(\mathbf{x})\sum_{k=1}^N w^k \delta\mathbf{w}^k,
\end{equation}
where the weights, $w$, and the first Brownian increments, $\delta\mathbf{w}$, of the $k^\text{th}$ sample trajectory are obtained from following stochastic simulation.
\begin{enumerate}
\item Set $i = 0,~\mathbf{X}_i = \mathbf{x}$.
\item $\mathbf{X}_{i+1} = \mathbf{X}_i + \mathbf{f}(\mathbf{X}_i)\delta t + B(\mathbf{X}_i)\mathbf{Z}_i\sqrt{\delta t}$, where $\mathbf{Z}_i\sim N(0,I_m)$.
\item If $\mathbf{X}_{i+1} \in D^\circ$, then $i=i+1$ and go to step 2.
\item If $\mathbf{X}_{i+1} \notin D^\circ$, then finish the simulation. Return $w^k = \exp\left(-\frac{1}{\lambda}(\phi(\mathbf{X}_{i+1})+\delta t\sum_{j=0}^iq(\mathbf{X}_j))\right)$ and $\delta\mathbf{w}^k=\mathbf{Z}_0$.
\end{enumerate}
Note that $\delta t$ is sufficiently small time step for simulation of a continuous stochastic process.

\subsection{Change of Measure} \label{subsec:ch_mea}
In the naive MC sampling process, the sample trajectories for the estimation are collected from the passive diffusion dynamics (\ref{eq:dyn_passive}).
Most trajectories, however, may be useless (i.e. they hit the obstacle or reach the goal region through very awkward way, which are far from optimum).
Rather than using naive MC sampling, it is possible to utilize advanced sampling technique to improve the quality of samples;
the importance sampling scheme is widely adopted in the path integral control literature.
Consider the new stochastic dynamics which drifts by some predefined deterministic process $\mathbf{u}_{\text{in}}$,
\begin{equation}
d\mathbf{x} = f(\mathbf{x})dt + G(\mathbf{x})\mathbf{u}_{\text{in}}dt + B(\mathbf{x})d\mathbf{w}^{(1)}, \label{eq:dyn_control2}
\end{equation}
and let $Q$ be the corresponding probability measure.

Then, the trajectories from the above stochastic dynamics can be used to estimate the desirability function and the optimal control, which is referred as a measure change or importance sampling.
Rewriting (\ref{eq:est_cost}) and (\ref{eq:est_cont}) yields
\begin{equation}
\psi(\mathbf{x}) = \int W(\vec{\mathbf{x}}) \frac{dP(\vec{\mathbf{x}})}{dQ(\vec{\mathbf{x}})}dQ(\vec{\mathbf{x}}) = E_Q\left[W(\vec{\mathbf{x}})\frac{dP(\vec{\mathbf{x}})}{dQ(\vec{\mathbf{x}})}\right], \label{eq:impor_sam1}
\end{equation}
and
\begin{equation}
\mathbf{u}^*(\mathbf{x})dt = \frac{1}{\psi(\mathbf{x})}G_c^{-1}(\mathbf{x})E_Q\left[W(\vec{\mathbf{x}})\mu(\mathbf{x})dt\frac{dP(\vec{\mathbf{x}})}{dQ(\vec{\mathbf{x}})}\right]. \label{eq:impor_sam2}
\end{equation}

The Radon-Nikodym derivative of $P$ with respect to $Q$, $\frac{dP(\vec{\mathbf{x}})}{dQ(\vec{\mathbf{x}})}$, can be obtained from following corollary.

\begin{corollary}[Girsanov's Theorem \cite{gardiner1985handbook}]
Suppose $P$ and $Q$ is the probability measures induced by the trajectories (\ref{eq:dyn_passive}) and (\ref{eq:dyn_control2}), respectively.
Then the Radon-Nikodym derivative of $P$ with respect to $Q$, $\frac{dP(\vec{\mathbf{x}})}{dQ(\vec{\mathbf{x}})}$, is given by
\begin{align}
&\frac{dP}{dQ} = \exp(-\frac{1}{2\lambda}\int^{t_f}_t\mathbf{u}_{\text{in}}'G_c'\Sigma_c^{-1}G_c\mathbf{u}_{\text{in}}dt \nonumber \\
&~~~~~~~~~~~~~~~~~~~~~~~~~-\frac{1}{\lambda} \int^{t_f}_t\mathbf{u}_{\text{in}}'G_c'\Sigma_c^{-1}B_cd\mathbf{w}^{(1)})\nonumber\\
&= \exp(-\frac{1}{2\lambda}\int^{t_f}_t\mathbf{u}_{\text{in}}'R\mathbf{u}_{\text{in}}dt -\frac{1}{\lambda} \int^{t_f}_t\mathbf{u}_{\text{in}}'G_c'\Sigma_c^{-1}B_cd\mathbf{w}^{(1)}).
\end{align}
\end{corollary}
With new probability measure $Q$, sampling procedure is changed as
\begin{enumerate}
\item Set $i = 0,~\mathbf{X}_i = \mathbf{x}$.
\item $\mathbf{X}_{i+1} = \mathbf{X}_i + \mathbf{f}(\mathbf{X}_i)\delta t + G(\mathbf{X}_i)\mathbf{u}_{\text{in}}(t_i)\delta t + B(\mathbf{X}_i)\mathbf{Z}_i\sqrt{\delta t}$, where $\mathbf{Z}_i\sim N(0,I_m)$.
\item If $\mathbf{X}_{i+1} \in D^\circ$, then $i=i+1$ and go to step 2.
\item If $\mathbf{X}_{i+1} \notin D^\circ$, then finish the simulation. Return $w^k = \exp\left(-\frac{1}{\lambda}(\phi(\mathbf{X}_{i+1})+\delta t\sum_{j=0}^iL_j)\right)$ and $\delta\mathbf{w}^k=\mathbf{Z}_0$, where $L_j \equiv q(\mathbf{X}_j)+\frac{1}{2}\mathbf{u}_{\text{in}}(t_j)'R\mathbf{u}_{\text{in}}(t_j)+\mathbf{u}_{\text{in}}(t_j)'G_c'\Sigma_c^{-1}B_c\mathbf{Z}_j/\sqrt{\delta t}$.
\end{enumerate}
The estimation of the desirability function is the same as (\ref{eq:est_cost2}) but because $\mu(\mathbf{x})dt = G_c(\mathbf{x})\mathbf{u}_{\text{in}}dt + B_c(\mathbf{x})d\mathbf{w}^{(1)}$ by substituting it to (\ref{eq:impor_sam2}), the estimation of the optimal control is given as,
\begin{equation}
\hat{\mathbf{u}}(\mathbf{x})\delta t = \mathbf{u}_{\text{in}}\delta t + \frac{1}{N\hat{\psi}(\mathbf{x})}G_c^{-1}(\mathbf{x})B_c(\mathbf{x})\sum_{k=1}^N w^k \delta\mathbf{w}^k. \label{eq:impor_con}
\end{equation}
Note that all the estimations are unbiased\cite{kappen2015adaptive,thijssen2015path}.
Especially, it is proven that the variance of estimation decreases as  $\mathbf{u}_{\text{in}}$ becomes closer to the real optimal control $\mathbf{u}^*$ \cite{thijssen2015path}.

\section{Sampling with Topology-Embedded Planner}
One can view that by using importance sampling, sample trajectories are obtained around (or biased to) the \textit{reference trajectory} induced by deterministic dynamics with $\mathbf{u}_{\text{in}}$.
Then through path integral procedure, the optimal trajectory/control is obtained by \textit{modifying} the reference trajectory/control.
One candidate of reference is the optimal trajectory for deterministic (noise-free) system.

However, the modification may force the result to local optimum if the amount of samples are not enough or the samples are far from global optimum.
Also, the problems addressed in this work may have many local optima, because the state space has high non-convexity due to obstacle regions.
The difficulty caused from non-convex space can be resolved if we have sample trajectories around every local optimum.

In this section, we propose the Path Integral with Rapidly-exploring Random Homology-embedded Tree (PI-RRHT*) algorithm in order to resolve such difficulty.
The algorithm consists of expansion (Algorithm \ref{alg:Expansion}) and execution (Algorithm \ref{alg:Execution}) phases,
where the former operates in lead-time and intermittently between execution phases, and latter runs on-line;
such construction has been utilized in \cite{jeon2015optimal}.
In expansion phase, the algorithm finds all different topological classes of trajectories for deterministic optimal motion planning problem.
And in execution phase, the control input for stochastic optimal control is computed in a receding horizon scheme with the path integral formula.

\subsection{Topological Representation of Trajectories in 2D}
Presence of obstacles in an environment differentiates topological classes among trajectories.
Suppose the configuration space, $\mathcal{C}$, and the obstacles are given by 2-dimensional subsets of $\mathbb{R}^2$.
Let $\sigma:[0,1]\rightarrow\mathbb{C}$ be a trajectory in the configuration space and $\sigma_1$ and $\sigma_2$ connecting the same start and end coordinates.
The two trajectories are called homologous if $\sigma_1$ together with $\sigma_2$ (the later with opposite orientation) forms the complete boundary of a 2-dimensional manifold embedded in $\mathcal{C}$ not containing/intersecting any of the obstacles \cite{bhattacharya2012topological}.

The configuration space can be represented as a subset of the complex plane $\mathbb{C}$, i.e. $(x,y)\in\mathcal{C} \Leftrightarrow x+iy\in\mathbb{C}$.
The obstacles are also represented as subsets of the complex plane, $\mathcal{O}_1, \mathcal{O}_2, ..., \mathcal{O}_N\subset\mathbb{C}$, and each obstacle has one \textit{representative point} which is denoted as $\zeta_l\in\mathcal{O}_l,~\forall l = 1,...,N$.
For a given set of representative points, the obstacle marker function, $\mathcal{F}:\mathbb{C}\rightarrow\mathbb{C}^N$, is defined as follows,
\begin{equation}
\mathcal{F}(z)=\left[\frac{1}{z-\zeta_1}, \frac{1}{z-\zeta_2}, \cdots, \frac{1}{z-\zeta_N}\right]'.
\end{equation}
Then, we can define \textit{$H$-signature}, $\mathcal{H}_2:C_1(\mathbb{C})\rightarrow\mathbb{C}^N$, which represent homology class of trajectory as:
\begin{equation}
\mathcal{H}_2(\sigma)=\frac{1}{2\pi}\textit{Im}\left(\int_\sigma\mathcal{F}(z)dz\right),
\end{equation}
where $C_1(\mathbb{C})$ is the set of all curves/trajectories in $\mathbb{C}$.

Especially, when the trajectory from $z_1$ to $z_2$ is \textit{short} enough (that is, a straight line connecting the same points is in same homology class), its $H$-signature can be calculated analytically as
\begin{equation}
\left(\mathcal{H}_2(e)\right)_l = \frac{1}{2\pi}\absmin_{k\in\mathbb{Z}}(\arg(z_2-\zeta_l)-\arg(z_1-\zeta_l)+2k\pi),
\end{equation}
where function $\absmin$ returns the value which have the minimum absolute value.

If two trajectories $\sigma_1$ and $\sigma_2$ connecting the same points have the same $H$-signatures, $\mathcal{H}_2(\sigma_1)=\mathcal{H}_2(\sigma_2)$, they are homologous and the reverse is also true.
Also, we can restrict the homology class of trajectories by defining disjoint sets of allowed and blocked $H$-signature, $\mathcal{A}$ and $\mathcal{B}$, where $\mathcal{U}=\mathcal{A}\cup\mathcal{B}$ and $\mathcal{U}$ denotes the set of the $H$-signatures of all trajectories.
By well restricting the allowed $H$-signature set, the topological motion planning algorithm can secure scalability with the number of obstacles.
It can be observed from Fig. \ref{fig:c00} and Fig. \ref{fig:c02} that there are plenty of trajectories which connect the same points and have different $H$-signatures.

The $H$-signatures for a higher dimensional space can be constructed by defining it directly \cite{bhattacharya2013invariants} or by using the configuration space mappings to 2-dimensional spaces \cite{Pokorny2016high}.

\subsection{Expansion phase of PI-RRHT*: Topology-Embedded Sampling-Based Planner}
\begin{algorithm}
\caption{PI-RRHT* algorithm (Expansion)}\label{alg:Expansion}
\begin{algorithmic}[1]
\State $(V,E)\gets(\emptyset,\emptyset);$
\For{$i=1 \textbf{ to } N_{iter}$}
\State $x_{new}\gets \textsc{Sampling}();$
\If{$x_{new}\in \chi_{goal}$}
\State $v_{new}\gets \textsc{AddRoot}(x_{new});~E_{new}\gets\emptyset;$
\Else
\State $(v_{new},E_{new}) \gets \textsc{ChooseParent}(V,x_{new});$
\EndIf
\If{$\sim\text{isempty}(v_{new}.N)$}
\State $V \gets V \cup v_{new};~E\gets E \cup E_{new};$
\State $(V,E) \gets \textsc{Rewire}(V, E, x_{new});$
\EndIf
\EndFor
\end{algorithmic}
\end{algorithm}
\vspace*{-.25in}
\begin{algorithm}
\caption{$\textsc{ChooseParent}(V,x_{new})$}\label{alg:ChooseParent}
\begin{algorithmic}[1]
\State $E_{new}\gets \emptyset;~V_{near\_f}\gets \textsc{NearForward}(V,x_{new});$
\For{$v_{near\_f}\in V_{near\_f}$}
\State $e_{near\_f}\gets \textsc{TPBVP}(x_{new}, v_{near\_f}.x);$
\If{$\textsc{ObstacleFree}(e_{near\_f})$}
\State $E_{new} \gets E_{new} \cup e_{near\_f};$
\For {$n \in v_{near\_f}.N$}
\State $n_{new} \gets \textsc{Propagate}(e_{near\_f},n);$
\State \textsc{AppendNode}$(v(x_{new}),n_{new});$
\EndFor
\EndIf
\EndFor
\State \Return $(v(x_{new}),E_{new})$
\end{algorithmic}
\end{algorithm}
\begin{algorithm}
\caption{$\textsc{Rewire}(V, E, x_{new})$}\label{alg:Rewire}
\begin{algorithmic}[1]
\State $V_{near\_b}\gets \textsc{NearBackward}(V,x_{new});$
\For{$v_{near\_b}\in V_{near\_b}$}
\State $e_{near\_b}\gets \textsc{TPBVP}(v_{near\_b}.x, x_{new});$
\If{$\textsc{ObstacleFree}(e_{near\_b})$}
\State $E \gets E \cup e_{near\_b};$
\EndIf
\EndFor
\State $Q\gets v(x_{new}).N;$
\While{$Q\neq\emptyset$} \Comment{exhaustive search}
\State $n\gets \textsc{POP}(Q);$
\For {$v_{near\_b} \text{ of } v(n)$}
\State $n_{new} \gets \textsc{Propagate}(e_{near\_b},n);$
\If{\textsc{AppendNode}$(v_{near\_b},n_{new})$}
\State $Q\gets \textsc{InsertQ}(Q, n_{new});$
\EndIf
\EndFor
\EndWhile
\State \Return $(V,E)$
\end{algorithmic}
\end{algorithm}\vspace*{-.05in}
This subsection will be devoted to explain the expansion phase of PI-RRHT* algorithm which aims to find all optimal trajectories in different homology classes for deterministic approximation problem.
The algorithm constructs a graph on state space based on Rapidly-exploring Random Graph (RRG) algorithm \cite{karaman2011sampling} and search over the graph to project a tree into $H$-signature augmented space.
The graph in state space is defined by a set of vertices, $V$, and edges, $E$, where each vertex is composed of a state, $v.x$, and set of associated nodes $v.N$.
Each node $n\in v.N$ has its $H$-signature, $n.H$, a cost, $n.c$, and a parent node $n.\text{parent}$.

Some required functions are described as follows:
\textsc{Sampling}() function returns a random state from the state space.
\textsc{AddRoot}$(x_{new})$ creates root vertex as $v.x = x_{new},~v.n.c = 0,~v.n.parent = \emptyset$ and $v.n.H = \mathbf{1}+\mathcal{H}(e_{new})$ where $e_{new}$ is the straight line between $x_{new}$ and the \textit{goal representative point}.
\textsc{ObstacleFree}$(e)$ takes a trajectory $e$ as an argument and checks whether it lies in obstacle free region.
\textsc{NearForward}$(V,x)$ and \textsc{NearBackward}$(V,x)$ functions return $\mathcal{O}(\log|V|)$ number of vertices (see \cite{karaman2011sampling}) among the set of vertices, $V$, from and to $x$, respectively.
\textsc{TPBVP}$(x_1, x_2)$ returns the optimal trajectory from $x_1$ to $x_2$ without considering obstacles (i.e. the solution of two point boundary value problem), which can be implemented in various ways with respect to the system dynamics and cost~\cite{webb2013kinodynamic,ha2013successive,jeon2015optimal,karaman2010optimal}.
\textsc{Propagate}$(e,n)$ returns the new node, $n_{new}$, which is created by propagating $n$ through $e$; the new node is given as $n_{new}.H = n.H + \mathcal{H}(e)$, $n_{new}.c = n.c + \text{Cost}(e)$ and $n_{new}.\text{parent}=n$, where $\mathcal{H}(e)$ and $\text{Cost}(e)$ denotes $H$-signature and cost of trajectory, $e$, respectively.
\textsc{AppendNode}$(v,n_{new})$ checks if the new node is in the allowed homology class and imposes a partial ordering to the set of nodes in the vertex of the form:
\begin{equation}
n_a<n_b \Leftrightarrow (n_a.H = n_b.H)\wedge(n_a.c < n_b.c). \label{eq:partialordering}
\end{equation}
When above condition holds, $n_a$ is said to be \textit{dominated} by $n_b$, meaning that the paths to the root from $n_a$ and $n_b$ are homologous but the cost of path from $n_a$ is smaller.
This function takes a state vertex, $v$, and new node, $n_{new}$ as arguments and first checks if the new node is in blocked homology class, $n_{new}.H \in\mathcal{B}$, or dominated by any existing nodes at $v$.
Then, if it is blocked or dominated, the function returns failure.
If it is not, the function appends the new node, checks if it dominates any nodes at $v$ and prunes when necessary.
Finally, $\textsc{InsertQ}(Q, n_{new})$ adds a new node into a queue and prunes nodes there when necessary.

The algorithm operates as shown in Algorithm \ref{alg:Expansion}.
In the main loop, it first samples a new state and checks if it is in the goal region.
If so, a new vertex is created as a root of the tree and if not, the algorithm attempts to connect a new vertex to the graph in Algorithm \ref{alg:ChooseParent}; if such connection succeed, the new vertex will contain some nodes.
With the successfully created new vertex, the algorithm adds it and the new (forward) edges to the graph on line 10 and finds/adds new backward edges on line 1-7 in Algorithm \ref{alg:Rewire}.
These procedures are analogous to the RRG algorithm \cite{karaman2011sampling}, which guarantees to asymptotically contain all possible trajectories through the space.

In Algorithm \ref{alg:Rewire}, after adding all new edges, all nodes in the new vertex are added to the queue, $Q$, on line 8.
Then the queue is exhaustively searched using uniform cost search, like Dijkstra's algorithm on line 9-17.
These procedures make the graph project the tree into $H$-signature augmented space by propagating nodes in the queue and pruning with criteria in (\ref{eq:partialordering}).

\subsection{Execution Phase of PI-RRHT*: Receding Horizon Path Integral Control with Topology-Guided Path Sampling}
\begin{algorithm}
\caption{PI-RRHT* algorithm (Execution)}\label{alg:Execution}
\begin{algorithmic}[1]
\State Given the current state $\mathbf{x}_{cur}$ and the Tree $(V,E);$
\While{$\mathbf{x}_{cur}\notin \partial D$}
\State $\vec{\mathbf{U}}_{in}\gets\textsc{ExtractReference}(\mathbf{x}_{cur}, (V,E));$
\State $\hat{\mathbf{u}}(\mathbf{x}_{cur}) \gets\textsc{PathIntegral}(\mathbf{x}_{cur},\vec{\mathbf{U}}_{in});$
\State $\mathbf{x}_{cur} \gets \textsc{ApplyControl}(\mathbf{x}_{cur}, \hat{\mathbf{u}}(\mathbf{x}_{cur}));$
\EndWhile
\end{algorithmic}
\end{algorithm}\vspace*{-.25in}
\begin{algorithm}
\caption{ExtractReference($\mathbf{x}_{cur}, (V,E)$)}\label{alg:ExtractReference}
\begin{algorithmic}[1]
\State $(v_{new},E_{new}) \gets \textsc{ChooseParent}(V, \mathbf{x}_{cur});$
\State $V \gets V \cup v_{new};~E \gets E \cup E_{new};$
\State $\{\vec{\mathbf{u}}_{in}^{(h)},~h=1,2,...,H\}\gets$\textsc{ReconstructPath}($G\gets(V, E),v_{new}$);
\Comment look at its ancestry to find the paths (node$\rightarrow$parent$\rightarrow$parent$\rightarrow$parent..., etc)
\State \Return{$\vec{\mathbf{U}}_{in}\gets\{\vec{\mathbf{u}}_{\text{in}}^{(h)},~h=1,2,...,H\}$}
\end{algorithmic}
\end{algorithm}
The execution phase of PI-RRHT* presented in Algorithm \ref{alg:Execution} computes and executes the optimal control for stochastic problem in a receding horizon fashion.
It consists of three procedures:
$\textsc{ExtractReference}(\mathbf{x}_{cur}, (V,E))$ shown in Algorithm \ref{alg:ExtractReference} takes the current state $\mathbf{x}_{cur}$ and the tree $(V, E)$ constructed from Algorithm \ref{alg:Expansion} as arguments and returns all open-loop optimal control tapes for the allowed homology trajectories from $\mathbf{x}_{cur}$ to the roots of the tree in $\chi_{goal}$.

Then in $\textsc{PathIntegral}(\mathbf{x}_{cur},\vec{\mathbf{U}}_{in})$, trajectories are sampled around each homology class and the optimal control is computed.
Suppose there are $H$ number of stochastic dynamics (\ref{eq:dyn_control2}) controlled by $\mathbf{u}_{\text{in}}^{(h)}$ and $Q_h,~h=1,2,...,H$ are corresponding probability measures.
Equation (\ref{eq:impor_sam1}) and (\ref{eq:impor_sam2}) can be rewritten as
\begin{equation}
\psi(\mathbf{x}) = \frac{1}{H}\sum_{h=1}^{H}E_{Q_h}\left[W(\vec{\mathbf{x}})\frac{dP(\vec{\mathbf{x}})}{dQ_h(\vec{\mathbf{x}})}\right],
\end{equation}
and
\HLB{
\begin{equation}
\begin{split}
&\mathbf{u}^*(\mathbf{x})dt \\
&= \frac{1}{H}\sum_{h=1}^{H}\frac{1}{\psi(\mathbf{x})}G_c^{-1}(\mathbf{x})E_{Q_h}
 \left[W(\vec{\mathbf{x}})\mu_h(\mathbf{x})dt\frac{dP(\vec{\mathbf{x}})}{dQ_h(\vec{\mathbf{x}})}\right].
\end{split}
\end{equation}
}
Suppose we sample $N$ trajectories from each homology class, $h = 1, 2, ..., H$, by procedure described in section \ref{subsec:ch_mea}\footnote{When the sampling procedure is performed, the reference control tape $\mathbf{u}_{\text{in}}$ needs to be augmented by $\mathbf{0}$ after its time length, because the problem is the first-exit type, i.e. final time is not fixed.} and let the weights of $k^\text{th}$ sample trajectory in $h^\text{th}$ homotopy class be indexed by $w^{(k,h)}$.
Then we have
\begin{equation}
\hat{\psi}(\mathbf{x}) = \frac{1}{H}\sum_{h=1}^H\hat{\psi}^{(h)}(\mathbf{x}), \label{eq:est_val_h}
\end{equation}
and
\begin{equation}
\hat{\mathbf{u}}(\mathbf{x})\delta t = \frac{1}{H}\sum_{h=1}^H\hat{\mathbf{u}}^{(h)}(\mathbf{x})\delta t, \label{eq:est_con_h}
\end{equation}
where $\hat{\psi}^{(h)}(\mathbf{x})\equiv \frac{1}{N}\sum_{k=1}^Nw^{(k,h)}$ and
\HLB{
\begin{equation*}
\begin{split}
&\hat{\mathbf{u}}^{(h)}(\mathbf{x})\delta t \\
&\equiv\frac{1}{N\hat{\psi}(\mathbf{x})}\sum_{k=1}^N w^{(k,h)}\left(\mathbf{u}^{(h)}_{\text{in}}\delta t+G_c^{-1}(\mathbf{x})B_c(\mathbf{x})\delta\mathbf{w}^{(k,h)}\right).
\end{split}
\end{equation*}
}
Note that from the above equations, the optimal control is only computed at the current time and state.
However, if the control policy is restricted as the open loop formulation, the state dependence term is dropped and we can obtain the \textit{open loop control tape} by replacing $\delta\mathbf{w}^k=\mathbf{Z}_i,~\forall i=2,3,...$ (see \cite{thijssen2015path}).
As a result, $\textsc{PathIntegral}(\mathbf{x}_{cur},\vec{\mathbf{U}}_{in})$ procedure computes the open loop policy for one-period of receding horizon.
Such control is applied to the system for one-period by $\textsc{ApplyControl}(\mathbf{x}_{cur}, \hat{\mathbf{u}}(\mathbf{x}_{cur}))$, then the overall algorithm repeats again until the state reaches the boundary of the domain.

\section{Numerical Example}

\begin{figure}[t]
	\centering
	\subfigure[]{
		\includegraphics*[width=.47\columnwidth, viewport =94 20 330 270]{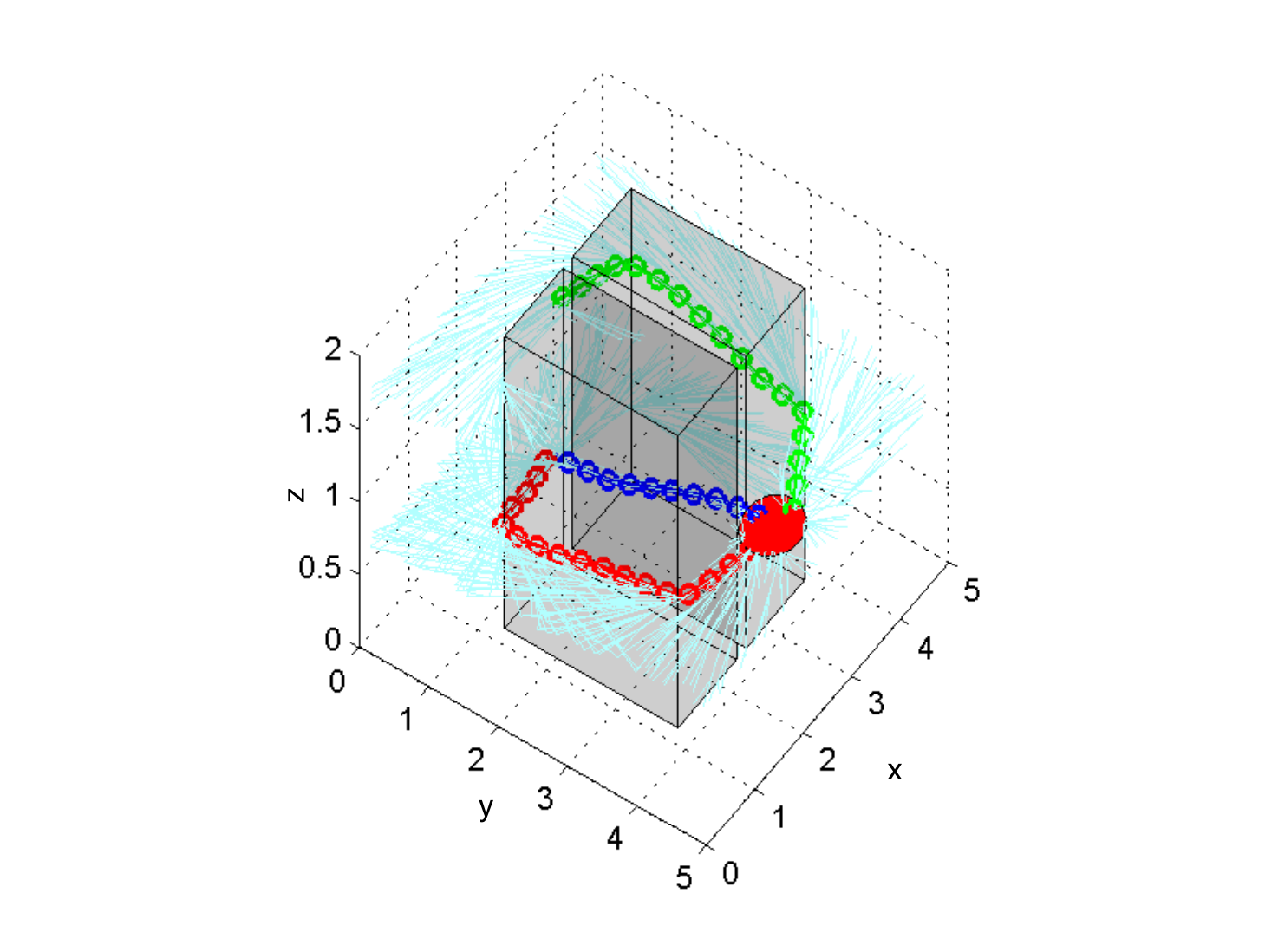}}
	\subfigure[]{
		\includegraphics*[width=.47\columnwidth, viewport =94 20 330 270]{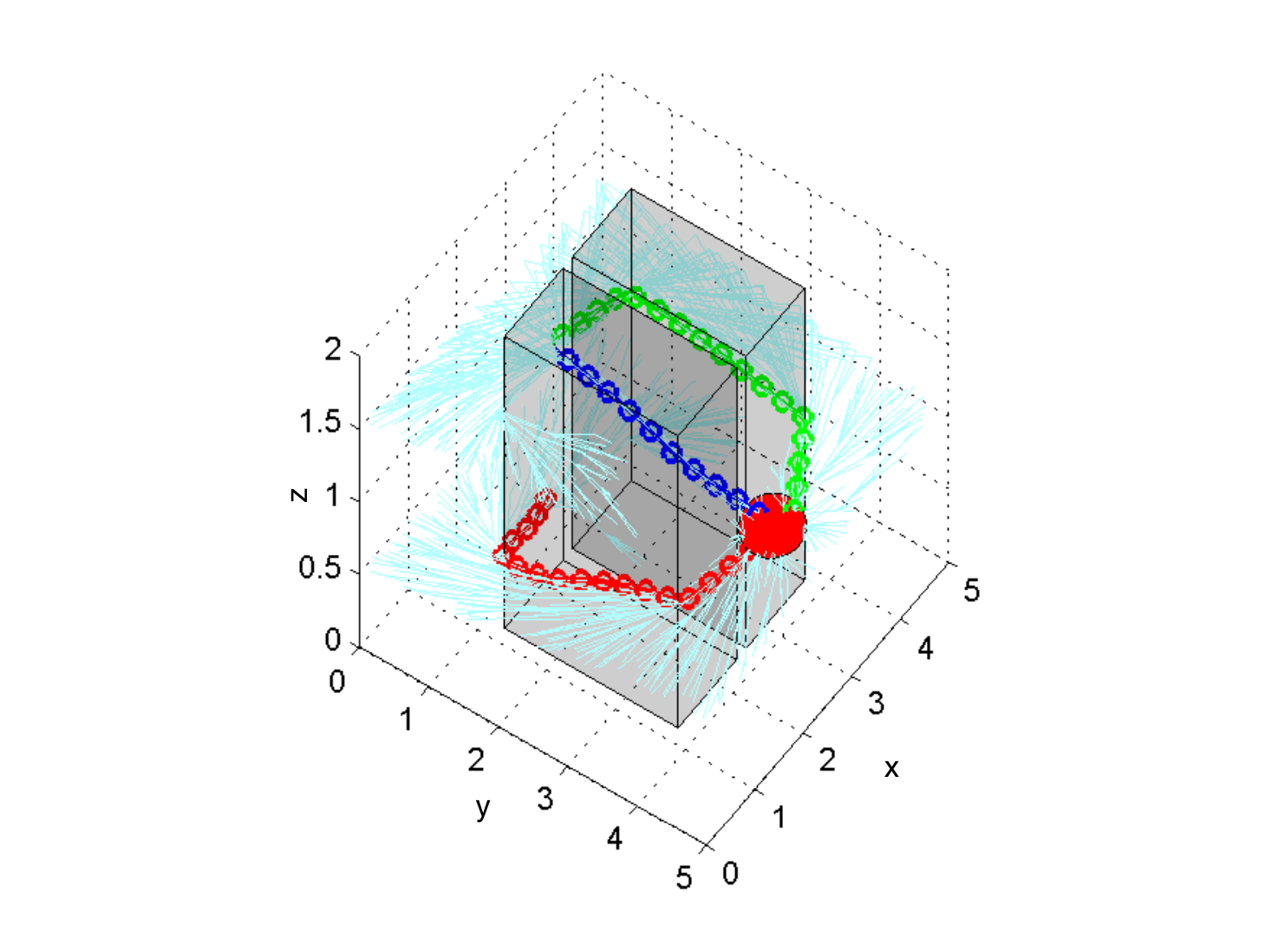}}
	\caption{The tree depicted by cyan edges is constructed by Expansion phase of PI-RRHT* (Algorithm \ref{alg:Expansion}) on $H$-signature augmented space for the single integrator example; a red circle represents goal region. $z$ axis denotes the values of $H$-signature with respect to the obstacle on (a) right side and (b) left side. The circled solid lines colored by red, blue and green result from \textsc{ExtractReference}() and show the deterministic optimal trajectories in different homology classes.}
	\label{fig:c00}
	\vspace*{-.15in}
\end{figure}
\begin{figure*}[t]
	\centering
	\subfigure[$b=0.1$]{
		\includegraphics*[width=4.2cm, height = 4.2cm, viewport =70 22 350 296]{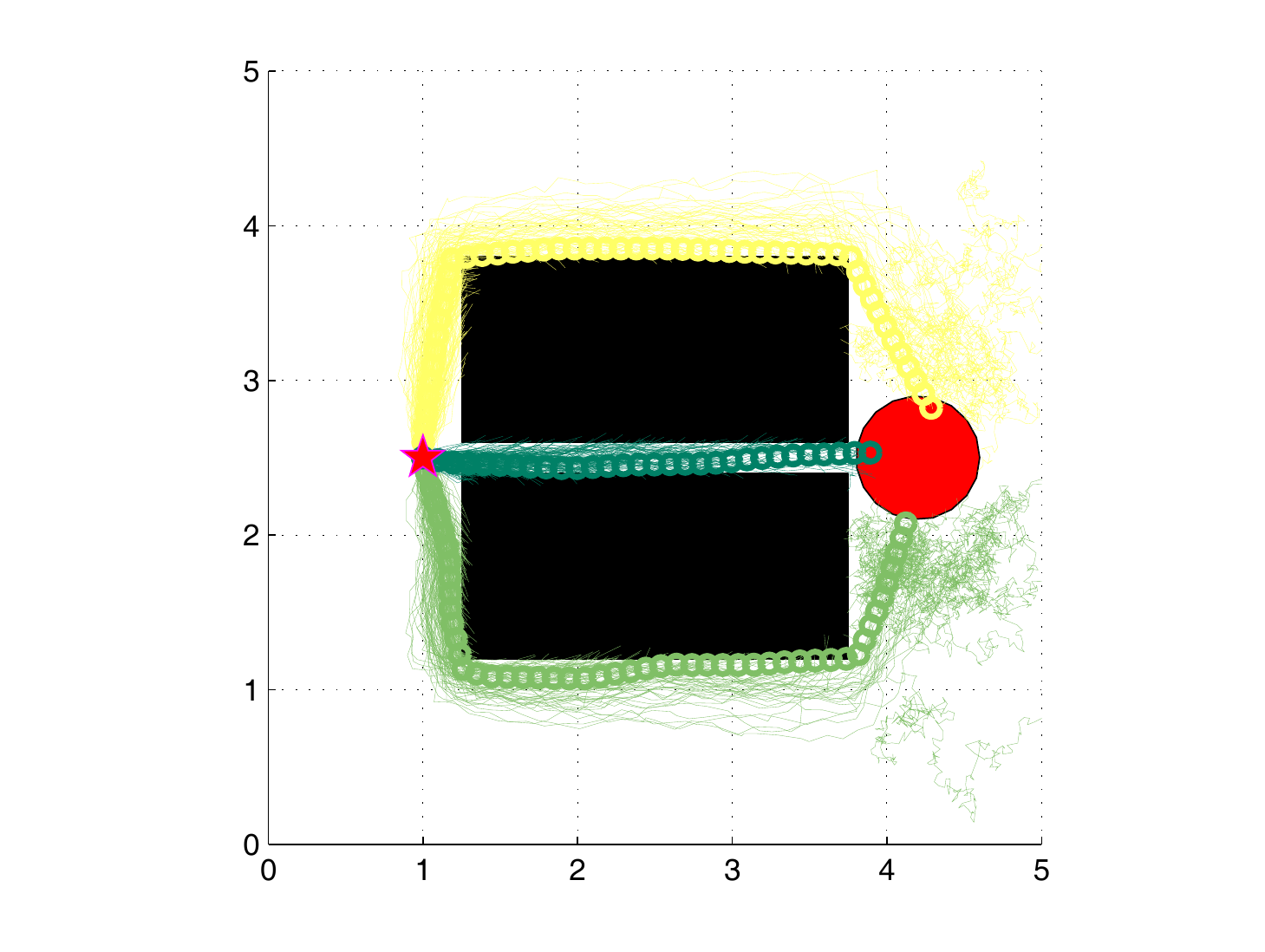}}
	\subfigure[$b=0.1$]{
		\includegraphics*[width=4.2cm, height = 4.2cm, viewport =70 22 350 296]{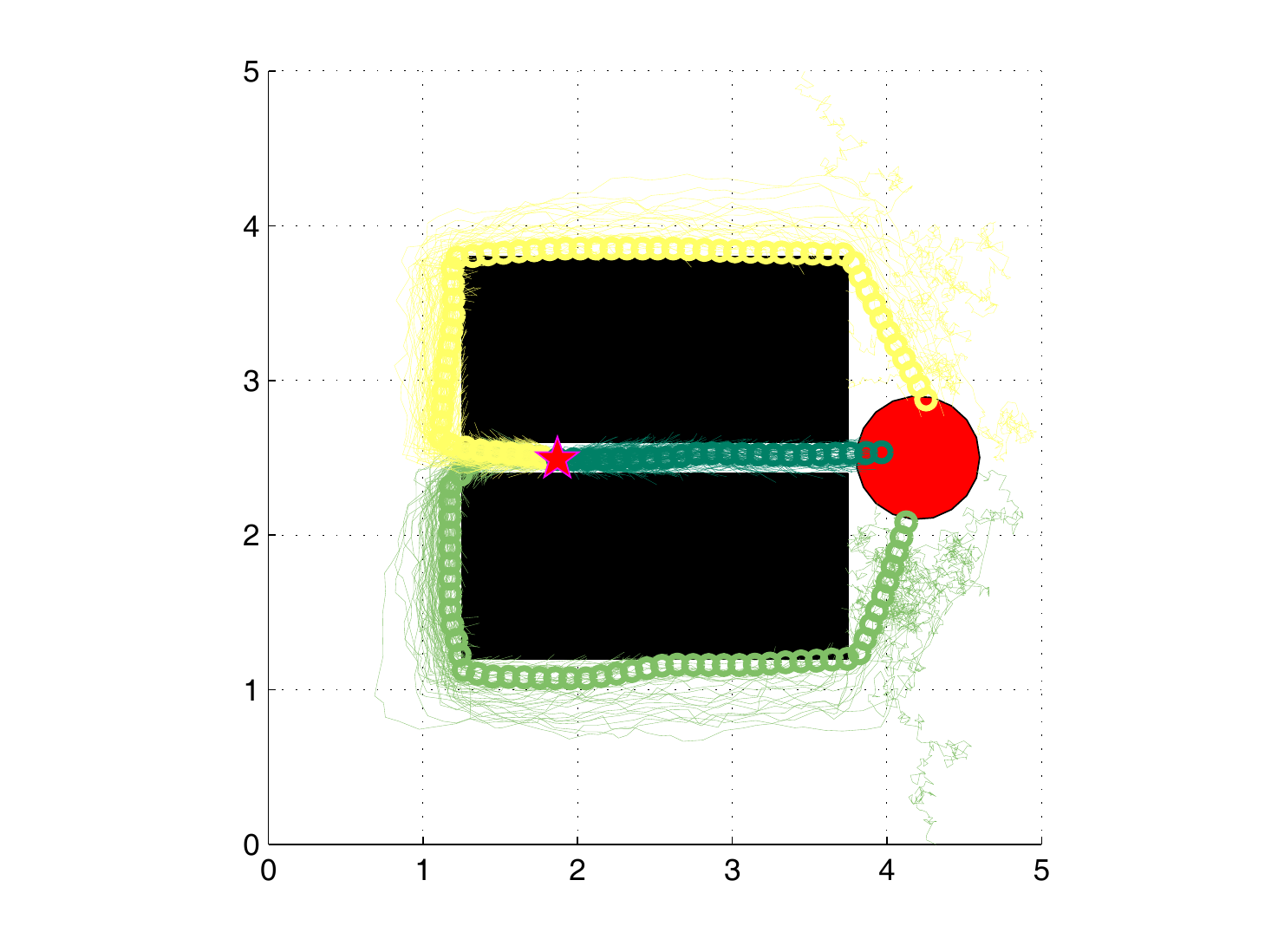}}
	\subfigure[$b=0.1$]{
		\includegraphics*[width=4.2cm, height = 4.2cm, viewport =70 22 350 296]{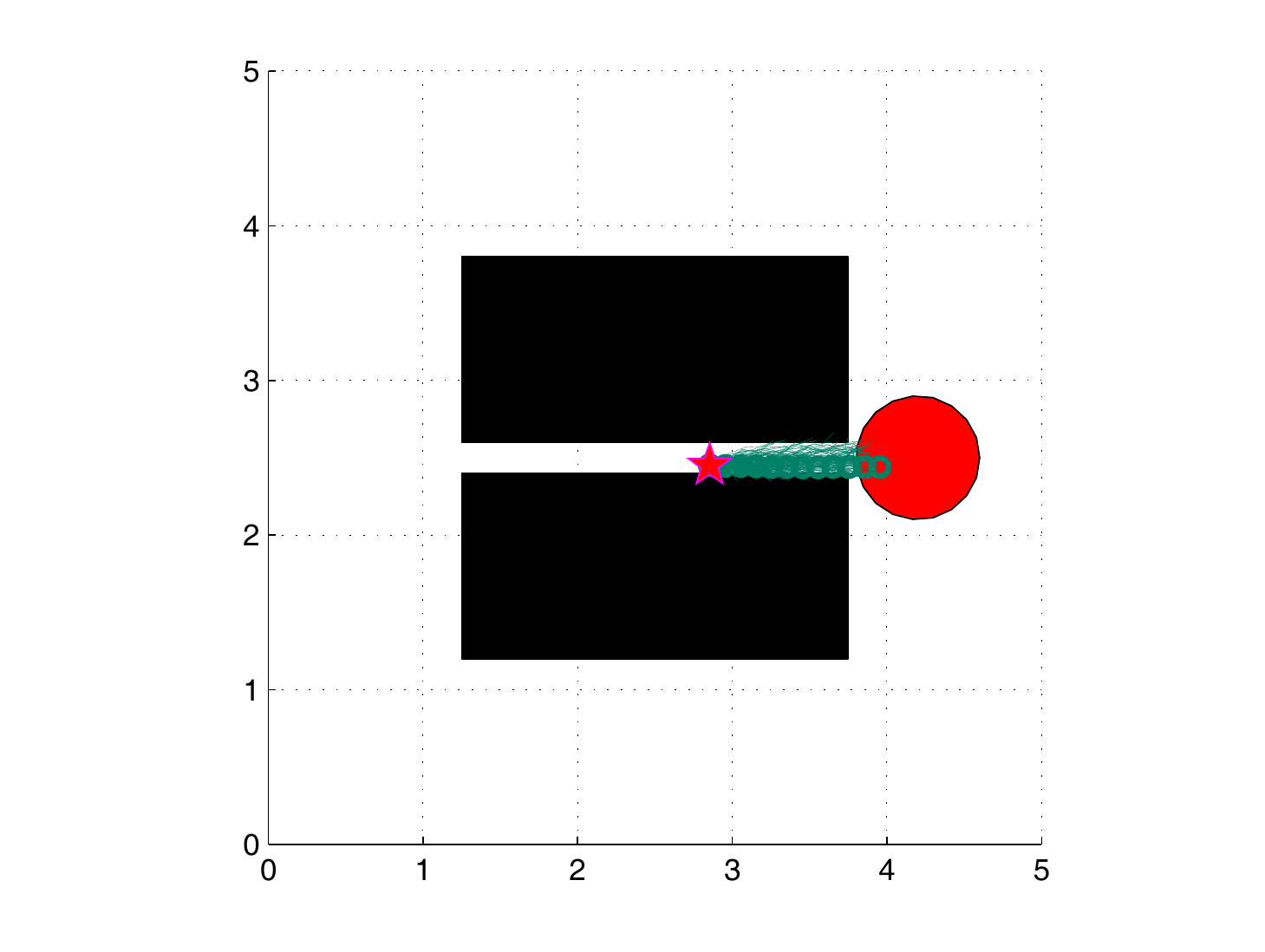}}
	\subfigure[$b=0.1$]{
		\includegraphics*[width=4.2cm, height = 4.2cm, viewport =70 22 350 296]{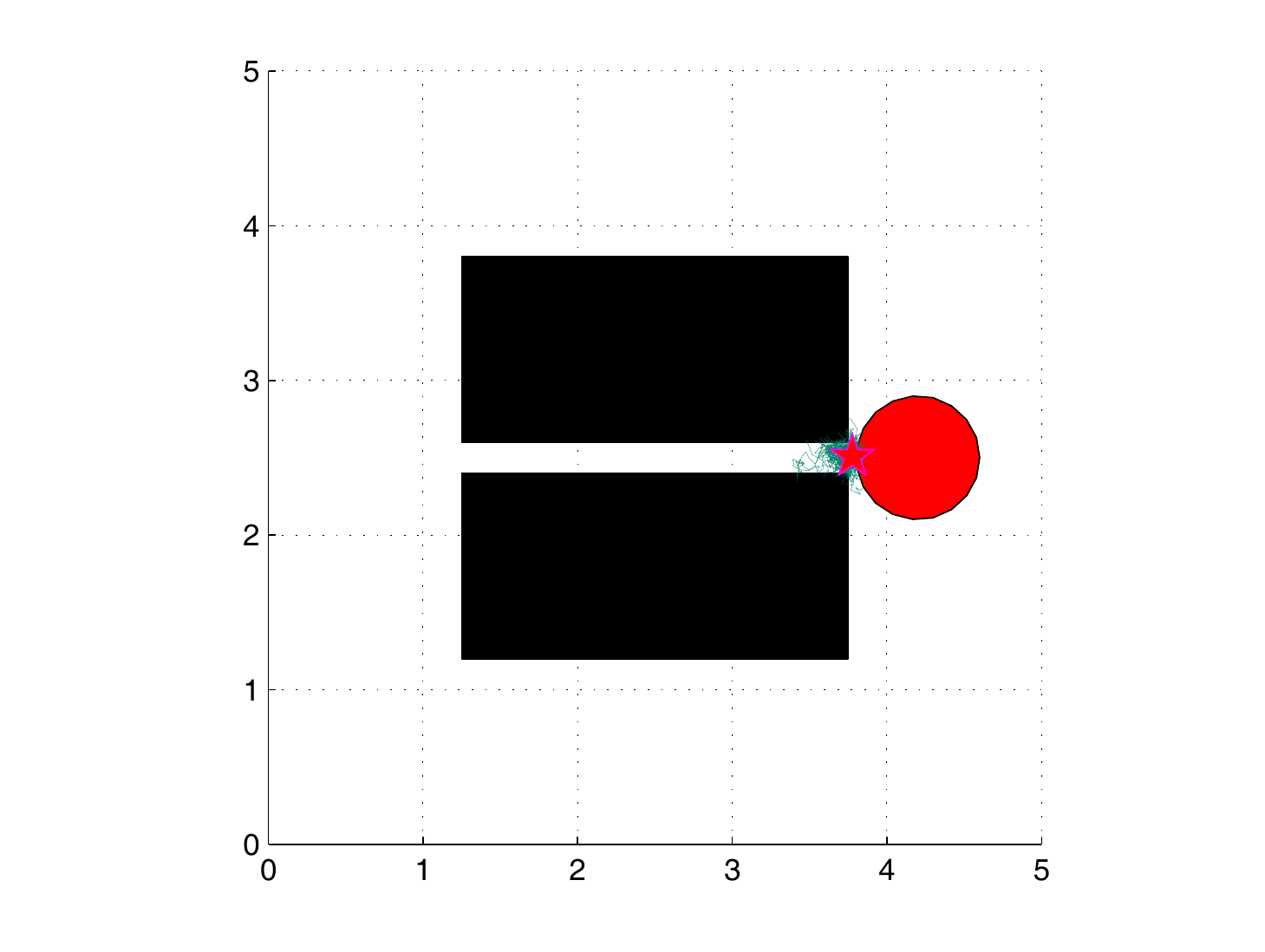}}
	\subfigure[$b=0.3$]{
		\includegraphics*[width=4.2cm, height = 4.2cm, viewport =70 22 350 296]{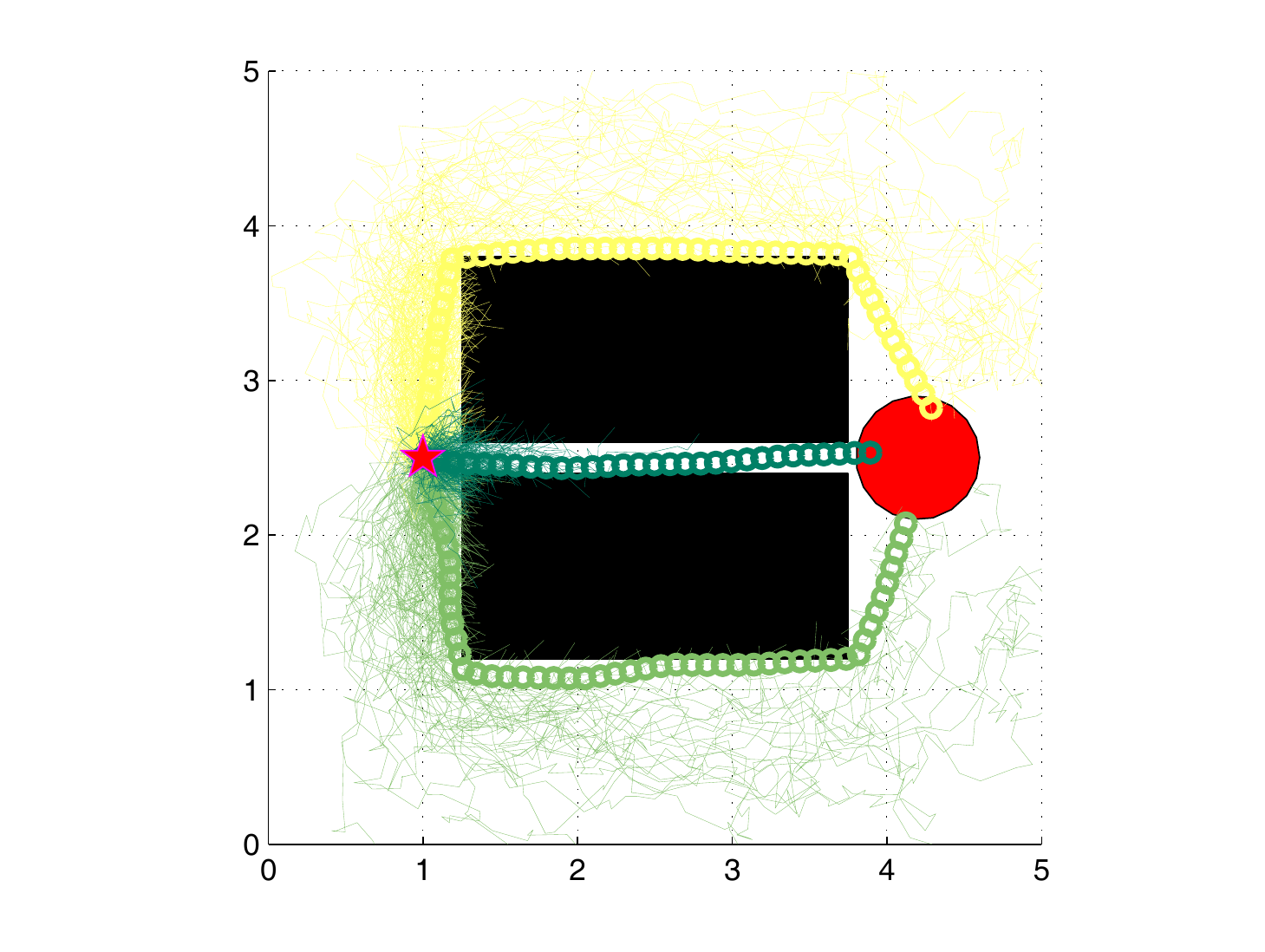}}
	\subfigure[$b=0.3$]{
		\includegraphics*[width=4.2cm, height = 4.2cm, viewport =70 22 350 296]{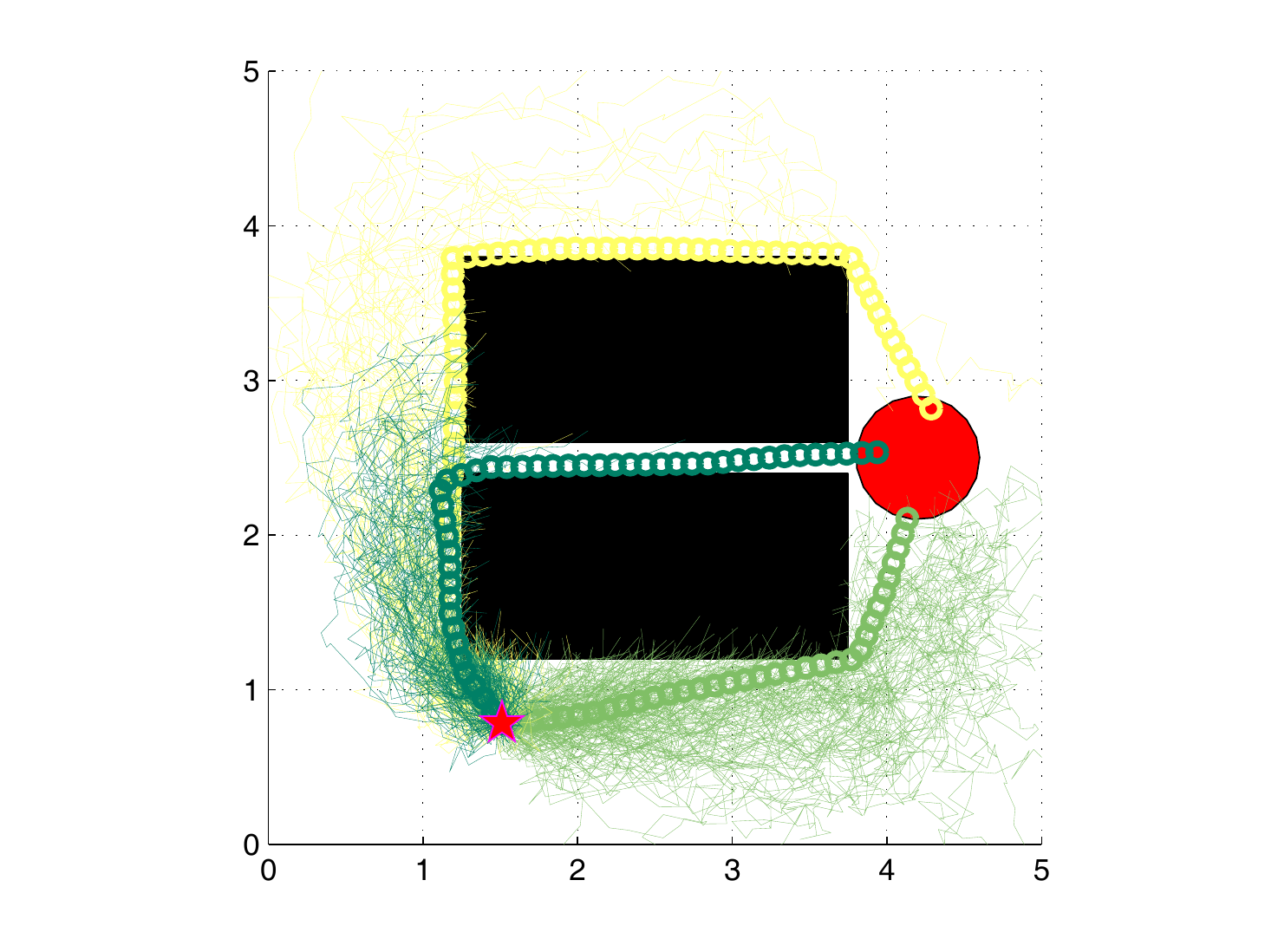}}
	\subfigure[$b=0.3$]{
		\includegraphics*[width=4.2cm, height = 4.2cm, viewport =70 22 350 296]{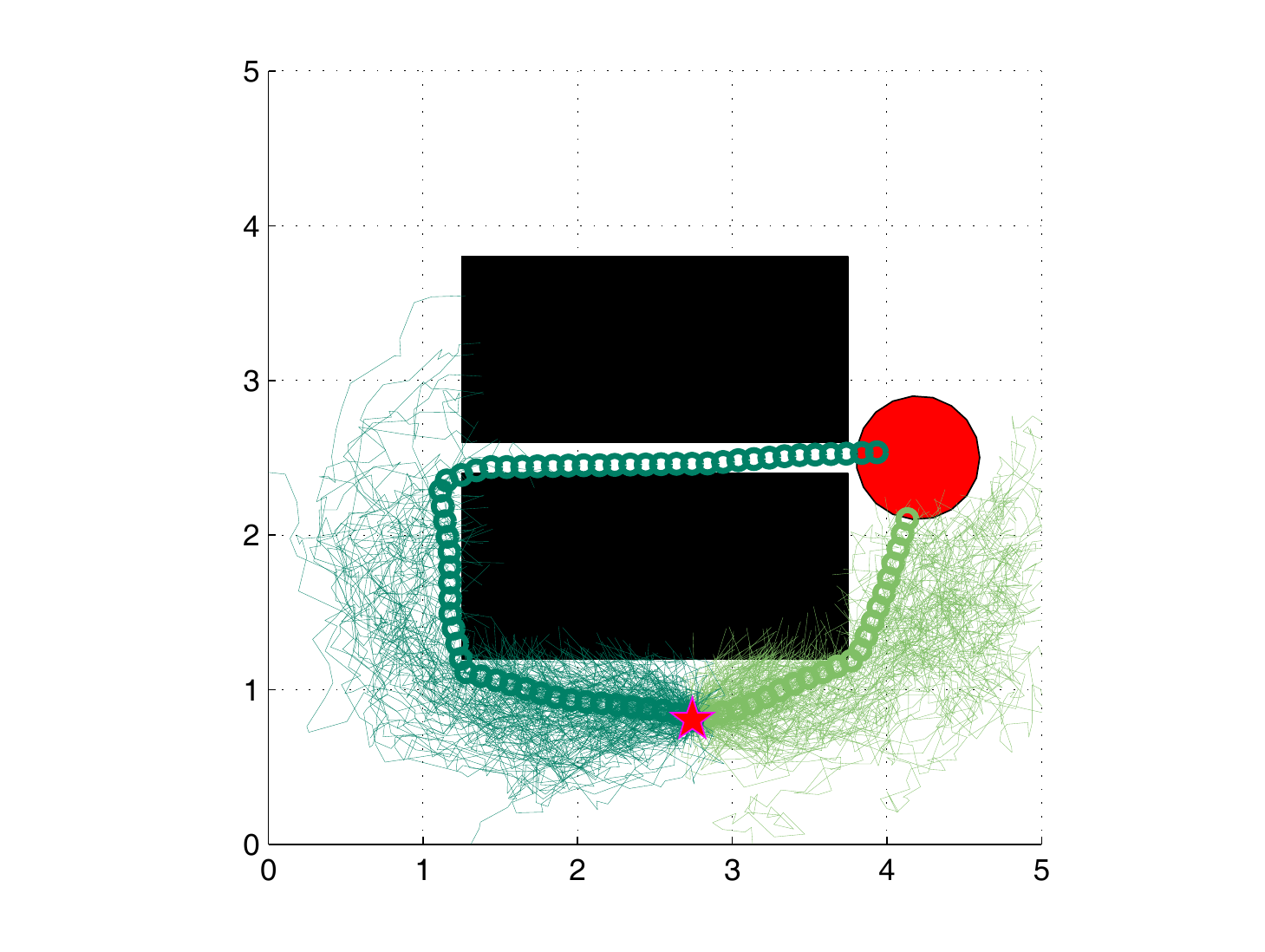}}
	\subfigure[$b=0.3$]{
		\includegraphics*[width=4.2cm, height = 4.2cm, viewport =70 22 350 296]{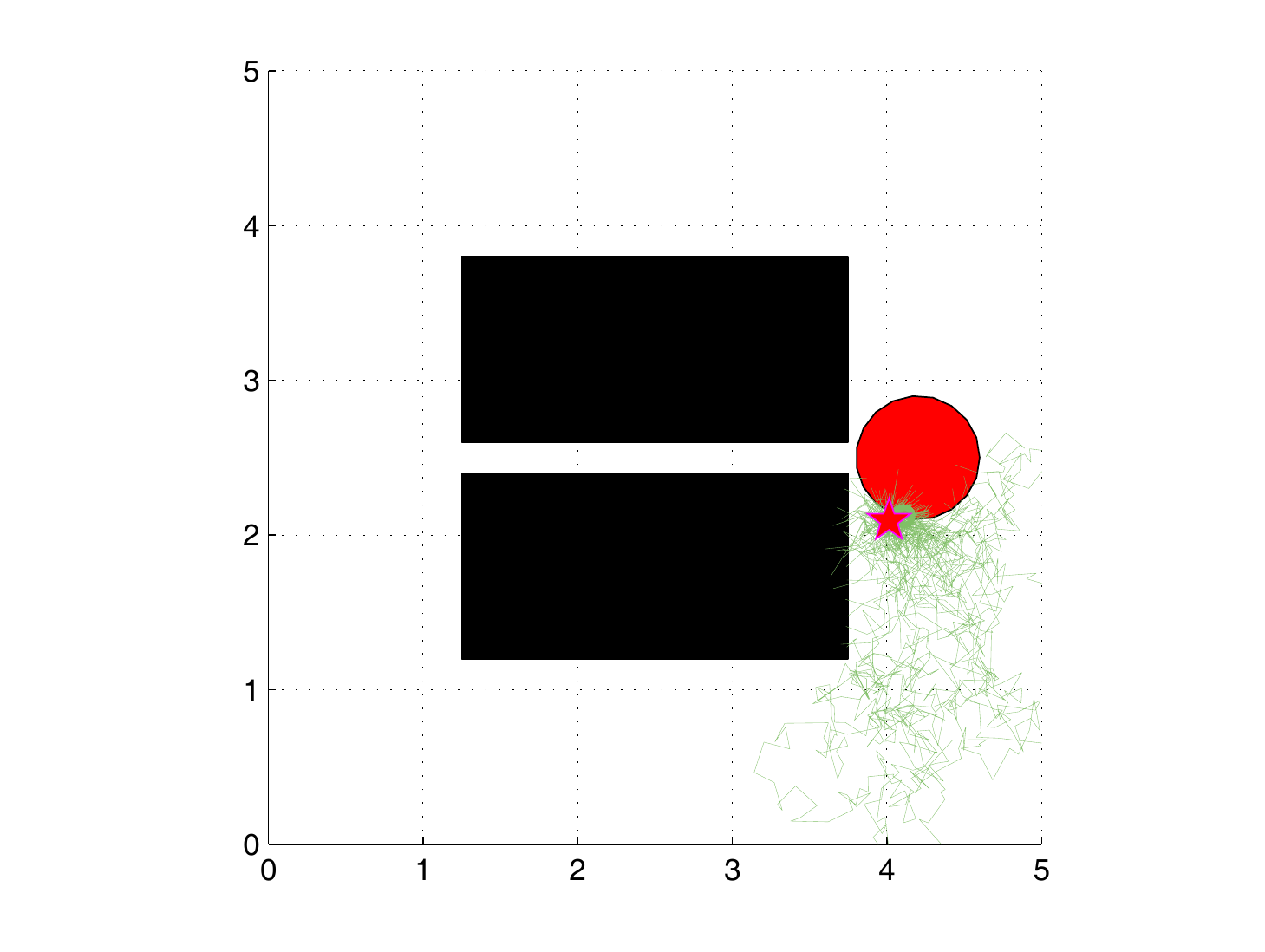}}
	\caption{Snapshots of execution phase of PI-RRHT* (Algorithm \ref{alg:Execution}) for the single integrator example, where $B=bI_2$ with (a)-(d) $b=0.1$ and (e)-(h) $b=0.3$. Colors of yellow, dark and bright green distinguish different homology classes, where thin edges and small-circled line represent the sample trajectories and the corresponding reference, respectively.}
	\label{fig:c01}
	\vspace*{-.15in}
\end{figure*}

\begin{figure}[t]
	\centering
	\subfigure[Path lengths vs. Iteration]{
		\includegraphics*[width=.62\columnwidth, viewport  = 40 20 400 300]{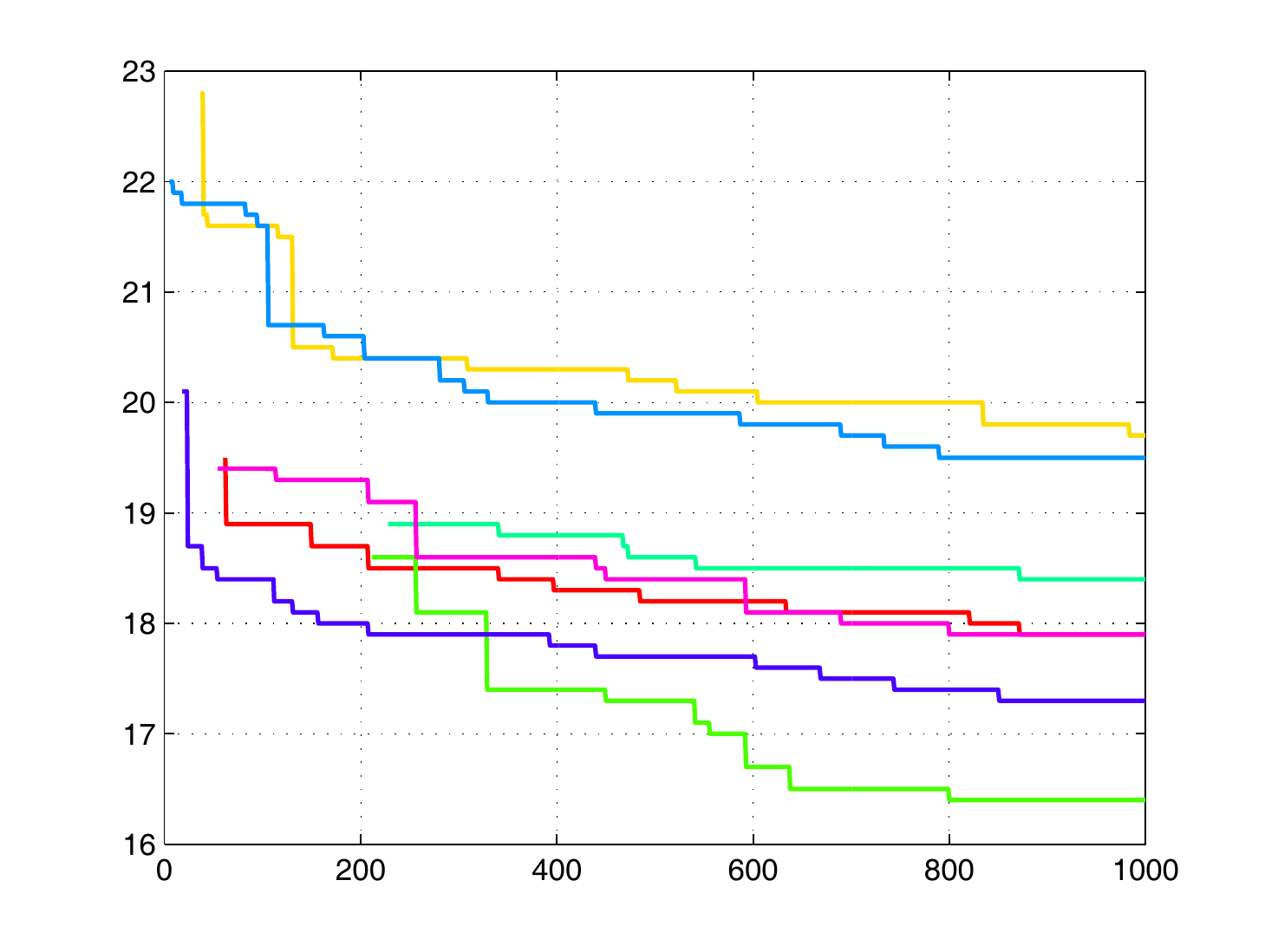}}
	\subfigure[]{
		\includegraphics*[width=.32\columnwidth, viewport =125 22 302 296]{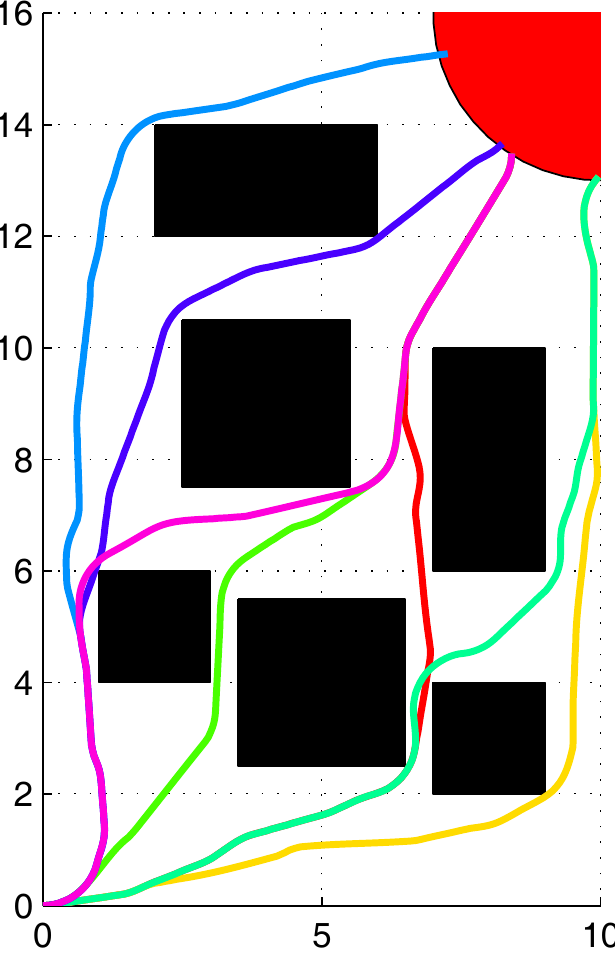}}
	\caption{Wheeled-mobile robot example. (a) Lengths of the shortest paths from $\mathbf{x}_{cur}=[0,0,0]'$ to goal region in each homology class at each iteration. (b) The resulting reference trajectories when 1000 vertices are added to the trees.}
	\label{fig:c02}
	\vspace*{-.15in}
\end{figure}

\begin{figure}[t]
	\centering
	\subfigure[]{
		\includegraphics*[width=.31\columnwidth, viewport =135 32 300 295]{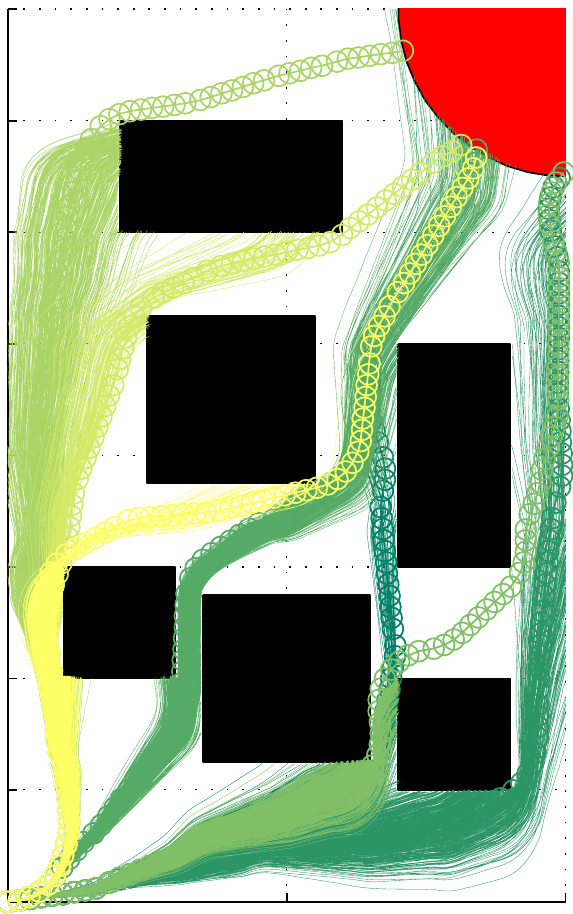}}
	\subfigure[]{
		\includegraphics*[width=.31\columnwidth, viewport =135 32 300 295]{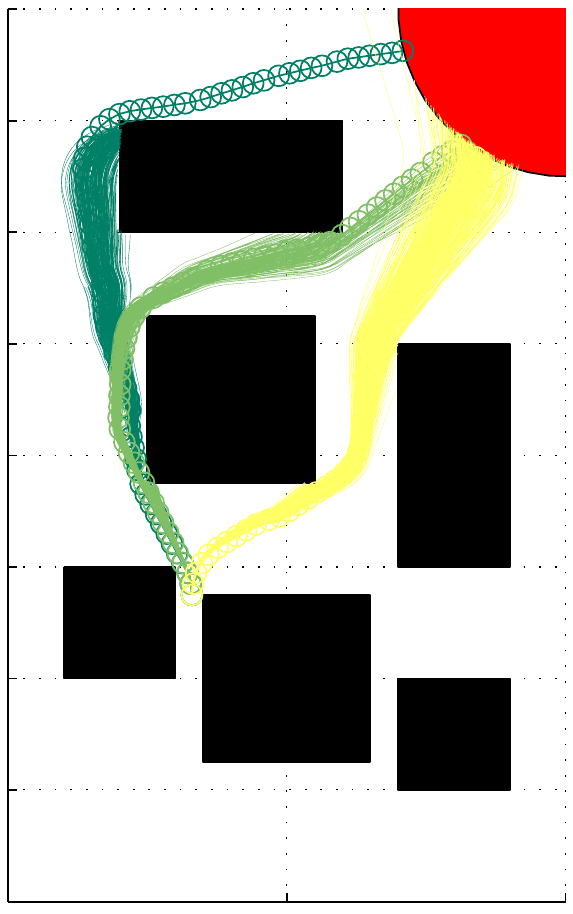}}
	\subfigure[]{
		\includegraphics*[width=.31\columnwidth, viewport =135 32 300 295]{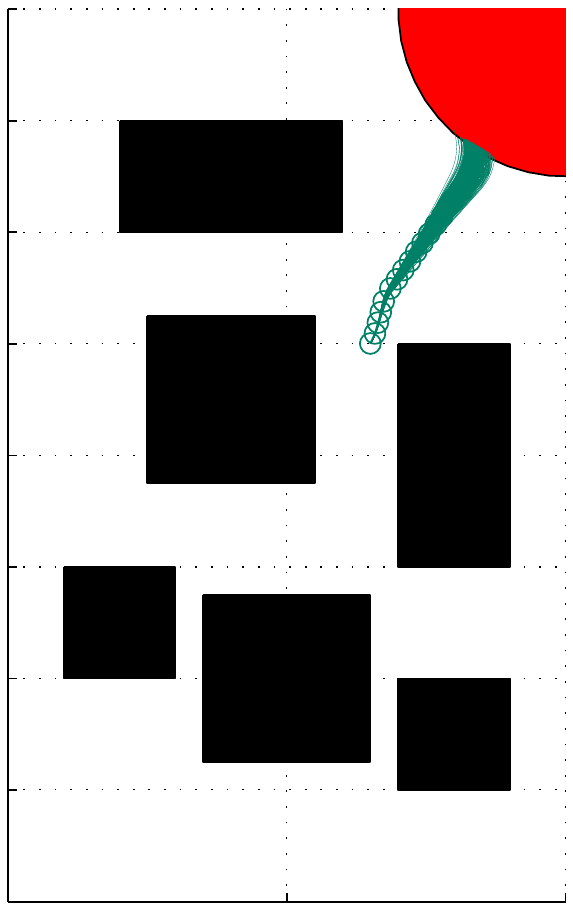}}
	\caption{Snapshots of the execution phase (Algorithm \ref{alg:Execution}) for the wheeled-mobile robot example.}
	\label{fig:c03}
	\vspace*{-.15in}
\end{figure}
For the first example, we consider a simple two-dimensional stochastic single integrator in the environment having two obstacles.
The dynamics and the cost rate are given by:
$$\mathbf{f}(\mathbf{x}) = \mathbf{0},~G(\mathbf{x}) = I_2,~q(\mathbf{x}) = 1\text{ and}~R(\mathbf{x}) = 2I_2,$$
i.e. the position of a  robot in the configuration space, $\mathbf{x}\in D\subset\mathbb{R}^2$, is controlled by the velocity input, $\mathbf{u}\in\mathbb{R}^2$, while the objective of control is to reach the goal region while minimizing the cost function, $J = E\left[\phi(\mathbf{x}(t_f))+\int_0^{t_f}1+\mathbf{u}'\mathbf{u}dt\right]$.
At the boundary, the final cost is given as:
\begin{equation}
\phi(\mathbf{x}) = \left\{ \begin{array}{ll}
         0 & \text{if}~\mathbf{x} \in \partial D_{goal},\\
	        \infty & \text{otherwise}.\end{array} \right.
\end{equation}

The state is driven also by a diffusion term that contains the 2-dimensional Brownian motion; two diffusion matrices are considered in this example for comparison:
$$
B(\mathbf{x}) = 0.1I_2,~0.3I_2.
$$
Through the path integral procedure, the time step for stochastic simulations and the number of samples for each reference trajectory are set as $\delta t = 0.1$ and $N = 200$, respectively.
Finally, the period of receding horizon control is given by $\delta t$.

Fig. \ref{fig:c00} shows the results of the expansion phase (Algorithm \ref{alg:Expansion}) and \textsc{ExtractReference} function (Algorithm \ref{alg:ExtractReference}) for $\mathbf{x}_{cur} = [1,2.5]'$.
It is observed that three trajectories in different homology classes are returned and they all connect the query state $\mathbf{x}_{cur}$ to the goal region.
When projecting the tree onto $H$-augmented space,  the set of allowable $H$-signature value is defined as $\mathcal{A} \equiv \{z: |1-z_i|\leq H_{limit},~i = 1,2\}$ with $H_{limit}=0.6$ to extract trajectories in physically meaningful homology classes; otherwise, infinitely many trajectories that include paths revolving around the obstacle could be obtained.

Fig. \ref{fig:c01} depicts snapshots of the receding-horizon control process of the execution phase (Algorithm \ref{alg:Execution})  with two different  diffusion matrices.
Note that with large degree of diffusion the effect of Brownian noise will be so critical that the robot cannot pass through the narrow slit between the obstacles.
It is observed from the figure that when the noise is not critical, the robot goes to the goal region directly but it makes a detour when the noise increases.
It can be seen that by considering topologically various trajectories as references, the path-integral formula provides not only computations of local optimum around each reference but also comparative advantages between references.

The second example deals with a stochastic wheeled-mobile robot dynamics in more cluttered environment.
The state $\mathbf{x}=[x,~y,~\theta]'$ represents the position and heading of a car, which is controlled by the turn rate, $u$.
The dynamics is given by:
$$\mathbf{f}(\mathbf{x}) = \begin{bmatrix} V\cos(\theta) \\ V\sin(\theta) \\ 0 \end{bmatrix},~G(\mathbf{x}) = \begin{bmatrix} 0 \\ 0 \\ 1/\rho \end{bmatrix}\text{ and}~B(\mathbf{x}) = \begin{bmatrix} 0 \\ 0 \\ b \end{bmatrix},$$
with $V=\rho=1$ and $b=0.03$, respectively.
The cost rate and the final cost are set as $q(\mathbf{x}) = R(\mathbf{x}) = 1$ and
$$\phi(\mathbf{x}) = \left\{ \begin{array}{ll}0 & \text{if}~\mathbf{x} \in \partial D_{goal},\\ 1000 & \text{otherwise}.\end{array} \right. $$
Finally, we set $N=300$ and $H_{limit}=0.8$.

In the expansion phase, TPBVP solver returns the minimum length Dubin's curve for the deterministic dynamics with $q=1,~R=0$ and $|u|\leq 1$.
Fig. \ref{fig:c02} shows the paths from $\mathbf{x}_{cur}=[0,0,0]'$ to goal region in each homology class when 1000 vertices are added to the trees and lengths of the paths at each iteration.
As it proceeds, the algorithm finds paths in various homology classes and refines those.
Fig. \ref{fig:c03} depicts some snapshots of the execution phase.
It is shown that, by using the diverse reference trajectories, the importance sampler can obtain sample trajectories exploring the entire state space to compute correct estimation of optimal control.

\section{Conclusions}
This paper has addressed a class of continuous-time, continuous-space stochastic optimal control problems on complex environment with obstacles.
A Feynman-Kac path integral formula and an importance sampling method have been presented for the first-exit time problem.
A topological concept embedded motion planner has been proposed to generate numerous reference trajectories in different homology classes.
Then we have proposed a receding-horizon scheme which samples the trajectories around each reference;
as a result, the proposed algorithm not only provides a dynamically feasible and collision-free trajectory but also effectively alleviates concern about local optima.
Numerical examples have demonstrated the validity of the proposed approach.

Note that in our scheme, a sample trajectory is obtained with \textit{open-loop reference control-tape}.
As a result, a sample trajectory diverges from the reference trajectory as time passes as shown in Fig. \ref{fig:c01} and \ref{fig:c03}.
It reduces sample efficiency;
if sample trajectories deviate a lot from the reference, they will make collision with high probability.
It is expected to be resolved by utilizing the parameterized (feedback) policy as a reference.
We leave it for a future work.




\section*{Acknowledgment}
This work was supported by Agency for Defense Development (under in part  contract  \#UD140053JD and in part contract \#UD150047JD).

\bibliographystyle{IEEEtran}
\bibliography{icra16}

\end{document}